\documentclass[10pt,conference]{IEEEtran}
\IEEEoverridecommandlockouts
\IEEEaftertitletext{\vspace{-1.25\baselineskip}}

\ifCLASSOPTIONcompsoc
  \usepackage[nocompress]{cite}
\else
  \usepackage{cite}
\fi

\usepackage{amsmath,amssymb,amsfonts}
\usepackage{graphicx}
\usepackage{textcomp}
\usepackage{xcolor}
\usepackage{varwidth}
\usepackage{microtype}
\usepackage{tikz}
\usepackage{qcircuit}
\usepackage{pgf}
\usepackage{physics}
\usepackage{bbm}
\usepackage{xcolor}
\usepackage{csquotes}
\usepackage{listings}
\usepackage{pgfplots}
\usepackage{url}

\definecolor{codegreen}{rgb}{0,0.6,0}
\definecolor{codegray}{rgb}{0.5,0.5,0.5}
\definecolor{codepurple}{rgb}{0.58,0,0.82}
\definecolor{shadecolor}{named}{lightgray}

\lstdefinestyle{mystyle}{
    backgroundcolor=\color{lightgray!30!white},   
    commentstyle=\color{codegreen},
    keywordstyle=\color{red!50!orange},
    numberstyle=\tiny\color{codegray},
    stringstyle=\color{codepurple},
    basicstyle=\linespread{0.95}\fontencoding{T1}\footnotesize\fontfamily{lmtt}\fontseries{c}\selectfont,
    breakatwhitespace=false,         
    breaklines=true,                 
    captionpos=b,                    
    keepspaces=true,                 
    numbers=left,                    
    numbersep=5pt,                  
    showspaces=false,                
    showstringspaces=false,
    showtabs=false,                  
    tabsize=2,
    columns=fullflexible,
    language=Python,
}
\usepackage{booktabs,chemformula}
\usepackage[shading]{moeptikz}
\moeptikzset{shading=false}
\tikzset{>=latex}
\usepackage{tabularx}

\usetikzlibrary{
    arrows,
    positioning,
    decorations.pathmorphing,
    decorations.markings,
    decorations.pathreplacing,
    shapes,
    fadings,
    calc
}

\usepackage{algorithm}
\usepackage[noend]{algpseudocode}
\algnewcommand\algorithmicforeach{\textbf{for each}}
\algdef{S}[FOR]{ForEach}[1]{\algorithmicforeach\ #1\ \algorithmicdo}
\algrenewcommand\algorithmicindent{0.55em}%

\newcounter{thm}
\newcounter{ex}

\newtheorem{definition}[thm]{Definition}
\newtheorem{example}[ex]{Example}

\def\BibTeX{{\rm B\kern-.05em{\sc i\kern-.025em b}\kern-.08em
    T\kern-.1667em\lower.7ex\hbox{E}\kern-.125emX}}
\pgfplotsset{compat=1.17} 
\begin{document}
\title{Quantum Algorithms and Simulation for Parallel and Distributed Quantum Computing}

\author{
	\IEEEauthorblockN{Rhea Parekh}
	\IEEEauthorblockA{
		\textit{Independent Scholar}\\
		\textit{rheaparekh12@gmail.com}
		}\and
	\IEEEauthorblockN{Andrea Ricciardi}
	\IEEEauthorblockA{\textit{Independent Scholar} \\
		\textit{andrea.ricciardi@live.com}\\
	}
	\and
	\IEEEauthorblockN{Ahmed Darwish, Stephen DiAdamo}
	\IEEEauthorblockA{
		\textit{Technische Universit\"at M\"unchen}\\
		\textit{\{a.darwish, stephen.diadamo\}@tum.de} 
	}
}
\maketitle

\begin{abstract}
		
	A viable approach for building large-scale quantum computers is to interlink small-scale quantum computers with a quantum network to create a larger  distributed quantum computer. When designing quantum algorithms for such a distributed quantum computer, one can make use of the added parallelization and distribution abilities inherent in the system. An added difficulty to then overcome for distributed quantum computing is that a complex control system to orchestrate the various components is required. In this work, we aim to address these issues. We explicitly define what it means for a quantum algorithm to be distributed and then present various quantum algorithms that fit the definition. We discuss potential benefits and propose a high-level scheme for controlling the system. With this, we present our software framework called Interlin-q, a simulation platform that aims to simplify designing and verifying parallel and distributed quantum algorithms. We demonstrate Interlin-q by implementing some of the discussed algorithms using Interlin-q and layout future steps for developing Interlin-q into a control system for distributed quantum computers.
		
\end{abstract}


\section{Introduction}

\IEEEPARstart{S}{caling} quantum computers up to levels where practical quantum algorithms can be executed will require a number of technological breakthroughs. In the present state of technology, scaling quantum computers past the 100 qubit mark has proven challenging \cite{preskill2018quantum}. Even when quantum computers can support a large number of qubits in a single system, if current methods error correction methods like surface codes are used, the amount of control signals required to perform error correction will scale with the number of qubits, potentially bottle-necking logical instructions for an algorithm's execution\cite{campbell2019applying}. To overcome these obstacles, a potential solution is to instead create smaller-scale quantum computers and interlink them using a quantum network to perform quantum algorithms over a distributed system. The benefit of using smaller, interlinked quantum processors is the ability to perform larger quantum circuits on more robust and controllable quantum processors albeit with the added---potentially easier---problem of using distribution methods. When one can use networked quantum computers, an additional ability to use parallelism in algorithm design is enabled.

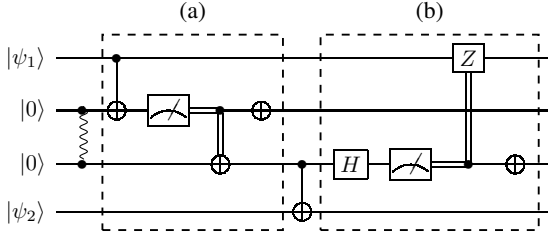
\begin{figure}
	\centering
	\begin{tikzpicture}[scale=1.0, every node/.style={transform shape}]
		\node (circ) at (0, 0) { 
			\scalebox{0.85}{
				\Qcircuit @C=1.0em @R=1.0em {
					\lstick{\ket{\psi_1}}    & \qw      & \ctrl{1} & \qw    & \qw       & \qw  & \qw     & \qw    & \qw  & \gate{Z} & \qw & \qw   \\
					\lstick{\ket{0}}         & \ctrl{0} & \targ    & \meter & \cctrl{1} & \targ  & \qw   &  \qw & \qw & \qw & \qw & \qw  \\
					\lstick{\ket{0}}         & \ctrl{0} & \qw      & \qw    & \targ     & \qw  &  \ctrl{1}   & \gate{H} & \meter & \cctrl{-2} & \targ & \qw \\
					\lstick{\ket{\psi_2}}    & \qw      & \qw      & \qw    & \qw       &\qw   & \targ     & \qw & \qw & \qw & \qw & \qw \\
				}
		}};
		\draw [dashed, line width=0.25mm] (-2.6, 1.3) rectangle (-0.2, -1.3);
		\draw [dashed, line width=0.25mm] (0.3, 1.3) rectangle (3.2, -1.3);
		\draw[decorate, decoration={snake, amplitude=0.5mm, segment length={1.3mm}}] (-2.85, -0.36) -- (-2.85, 0.35);
		\node[scale=0.9] at (-1.4, 1.6) {(a)};
		\node[scale=0.9] at (1.75, 1.6) {(b)};
	\end{tikzpicture}
	\caption{Circuit diagram for a non-local CNOT gate between $\ket{\psi_1}$ and $\ket{\psi_2}$ where (a) is the cat-entangler sequence and (b) the cat-disentangler sequence. The upper two qubits and the lower two qubits are physically separated between quantum computers.}
	\label{fig:cat-entangler}
\end{figure}

When moving from monolithic to distributed quantum computers, a variety of challenges arise. Indeed, there are many technological challenges to overcome towards building distributed quantum computers. A naturally arising problem to consider in this perspective is performing two-qubit operations between qubits that are physically separated between two quantum computers. To perform two-qubit operations with monolithic quantum technologies, generally the two qubits are physically near each other, and if not, swap-gates are applied to bring them near enough, known as the qubit routing problem \cite{cowtan2019qubit}. On the other hand, for two-qubit operations between distributed qubits, one needs a new technique for transporting the control information between devices. Possible options are to physically transmit qubits via a potentially noisy and lossy medium \cite{brown2016co}, using quantum teleportation \cite{zomorodi2018optimizing, van2007communication}, transferring control information to a flying qubit \cite{Daiss614, serafini2006distributed}, or using the method introduced in \cite{yimsiriwattana2004distributed} using one entangled pair and a two bits of classical communication as seen in Fig.~ \ref{fig:cat-entangler}.  

Once a method of performing non-local two qubit gates is selected, quantum circuits designed for monolithic systems need to then be remapped to a logically equivalent distributed version. To perform the remapping, one starts with the topology of the networked quantum computers, each with their own quantum processor chip structures. A monolithic circuit is converted such that any multi-qubit operation involving qubits located on different processors is replaced with a logically equivalent set of instructions orchestrating the additional tasks needed for the non-local operation. This remapping problem has been addressed in a variety of ways \cite{diadamo2020vqe, ferrari2020compiler, eisert2000optimal, daei2020optimized, andres2019automated, sundaram2021effic}, but until distributed quantum computing becomes more standardized, the most applicable method for generating and optimizing distributed circuits remains an open problem. 

The next problem arising is how to design and develop a control system for a distributed system of quantum computers. Already a step in this direction is the concept of cloud quantum computing which takes user input---usually as a circuit---and a software layer converts the input into control instructions for a single quantum computer \cite{devitt2016performing, cross2018ibm}. The quantum computer performs the computation and the results are sent back to the user via a communication network. For a distributed system of quantum computers, additional network connections are needed between the quantum computers. Moreover, the connections cannot simply be classical channels, but quantum channels will be needed for either distributing entanglement or moving data-containing qubits. Networked control systems for classical distributed systems have been developed in various scenarios \cite{brodtkorb2010state}, for example in GPU clusters \cite{mittal2015survey}, but a key problem that is not as critical for classical systems for computing is that the quantum computers need to be highly time-synchronized to perform joint measurements, for one. It is therefore a unique problem to design networked control systems for distributed quantum computers. Proposals addressing such control systems are found in \cite{diadamo2020vqe, haner2021distributed}. 

Finally, once the ability to perform distributed quantum algorithms is enabled, one can then start to consider the various quantum algorithms that can benefit from being distributed and parallelized. Such examples have been considered such as of distributed Shor's algorithm \cite{yimsiriwattana2004distributed}, Quantum Phase Estimation (QPE) \cite{neumann2020imperfect}, and accelerated Variational Quantum Eigensolver (VQE) \cite{diadamo2020vqe}. Further, a mathematical framework for expressing and analyzing distributed quantum algorithms has been developed in \cite{ying2009algebraic}. Now that the hardware technology is beginning to catch up with the theory, a relatively open field remains is to better understand what advantages---especially while considering the cost of execution---there really are to gain when moving into a distributed setting. 

In this work, we investigate two angles for distributed quantum computing. We consider firstly a formalization of parallel and distributed quantum programs and consider a collection of quantum algorithms fitting this formalization. Next, we introduce a novel software simulation tool for simulating distributed quantum algorithms called Interlin-q. Interlin-q is a Python library built on top of QuNetSim \cite{diadamo2020qunetsim}---a quantum network simulator---which generates and simulates the control steps needed in an asynchronous setting to simulate distributed quantum algorithms. The overall goal of the platform is to provide a tool for validating algorithms for distributing quantum circuits and testing control systems. In addition, one can use Interlin-q to simulate parallel and distributed algorithms to then benchmark the approaches for their distribution and parallelization efficiency. In this work, we provide an overview of the software library in its current state and some demonstrations. Overall, interlinking quantum computers to perform distributed quantum algorithms will inevitably be an important part of quantum computing in the coming future. This work aims to shed light on the open problems and foreseeable benefits of distributed quantum computing, an increasingly important topic for the future of quantum computing.

\section{Monolithic to Distributed Algorithms}

To start our investigation of distributed quantum algorithms, we generalize the concept of mapping monolithic quantum algorithms to distributed quantum programs and scheduling them for execution. Executing a distributed quantum algorithm on a distributed quantum computer has a general preparation and execution stages: 1) Allocate logical qubits within the network of quantum computers; 2) Remap circuits for the possibly distributed qubit assignment; 3) Generate a schedule for the control operations; 4) Distribute and execute the schedule; and 5) Merge the outputs. Some quantum algorithms, of which we investigate in the next section, have a particular structure that allows them to gain a large ``horizontal" speedups when parallelized, where other quantum algorithms requiring many logical qubits can more readily be executed on nearer-term quantum computers via a distributed quantum computer. To model this staged process of preparation and execution, we start with a QPU structure as collection of integers $Q =[q_1,...,q_k]$ representing a network of $k$ QPUs where each QPU $i$ has $q_i\in\mathbb{N}$ logical qubits. In this model, it is implied that the quantum network topology is completely connected  entanglement units are created during runtime. With this, we define a quantum parallel program.

\begin{definition}[Parallel Program]
	A program $P$ is the instruction-set needed to perform a \textit{monolithic} execution of a quantum circuit including the logical circuit and the number of times to repeat the execution of the circuit. A schedule $S(i)$ is a mapping from an execution-round number $i$ to sets of integers, where $|S(i)|$ is always the number of QPUs in the network. The $k$-th set of $S(i)$ represents the programs $\mathcal{P}_i \subset \{P_j\}_{j=1}^n$, where there are $n$ programs total to run, executing at time $i$ on QPU $k$ where two distinct sets in $S(i)$ are not necessarily disjoint. A collection of programs $\{P_j\}_{j=1}^n$, a function $M: O^n \mapsto O$ for $O$ the output of a program which acts as a central merging function, and a schedule form a parallel program $\mathcal{P} = \{\{P_1,...,P_n\}, S(i), M\}$. 
\end{definition}

\begin{definition}[Distributed Program]
	Given QPUs $Q = [q_1, ..., q_k]$, a distributed program $dP$ is a program $P$ where the circuit execution instructions of $P$ are assigned to qubits from multiple distinct QPUs from $Q$. In this framework, it implies there exists an $i$ where there are at least two distinct sets both containing $P$.
\end{definition}

\begin{figure}
	\centering
	\begin{tikzpicture}[scale=0.85, every node/.style={transform shape}]
		\node (i) at (0, 1) {$i=0$};
		\node (init) at (0, 0) {$S(i)$};
		\node[draw,line width=.25mm] (m1) at (-1.75, -1.75) {$QPU_1$};
		\node[draw,line width=.25mm] (m2) at (-0.25, -1.75) {$QPU_2$};
		\node () at (0.75, -1.75) {$\dots$};
		\node[draw,line width=.25mm] (mk) at (1.75, -1.75) {$QPU_k$};
		\node[draw,line width=.25mm] (gamma_vec) at (0, -3.5) {$\hat{\gamma}$};
		\node[draw,line width=.25mm] (merge) at (0, -4.60) {$M(\gamma_{1}, ..., \gamma_{n})$};
		\node () at (-1.45, -0.9) {$dP_{s_1^{(i)}}$};
		\node () at (0.35, -0.9) {$dP_{s_2^{(i)}}$};
		\node () at (1.45, -0.9) {$dP_{s_k^{(i)}}$};
		\node () at (3.85, -1.65) {$i+1$};
		\node () at (1.75, -3.8) {$i \leq r$};
		\node () at (-1.35, -2.65) {$\gamma_{s_1^{(i)}}$};
		\node () at (0.27, -2.65) {$\gamma_{s_2^{(i)}}$};
		\node () at (1.40, -2.65) {$\gamma_{s_k^{(i)}}$};
		        
		\draw[->,line width=.25mm] (i) -- (init);
		\draw[->,line width=.25mm] (init) -- (m1);
		\draw[->,line width=.25mm] (init) -- (m2);
		\draw[->,line width=.25mm] (init) -- (mk);
		\draw[->,line width=.25mm] (m1) -- (gamma_vec);
		\draw[->,line width=.25mm] (m2) -- (gamma_vec);
		\draw[->,line width=.25mm] (mk) -- (gamma_vec);
		\draw[->,line width=.25mm] (gamma_vec.south) -- (merge.north);
		        
		\draw[-,line width=.25mm] (gamma_vec.east) -- ([xshift=3cm]gamma_vec.east);
		\draw[-,line width=.25mm] ([xshift=3cm]gamma_vec.east) -- ([xshift=3cm, yshift=3.5cm]gamma_vec.east);
		\draw[->,line width=.25mm] ([xshift=3cm, yshift=3.5cm]gamma_vec.east) -- (init.east);
		\draw[->,line width=.25mm] (merge.south) -- ([yshift=-4mm]merge.south);
		        
	\end{tikzpicture}
	\caption{A depiction executing a parallel program. The system starts at time $i=0$, loading the programs specified by $S(i)$ to the respective QPUs until all $r$ rounds are ran. The outputs of the distributed programs are accumulated in an output vector $\hat{\gamma}$ during execution. Finally $M$ maps the collection of $n$ outputs $\hat{\gamma}$ to a single output.}
	\label{fig:parallel}
\end{figure}
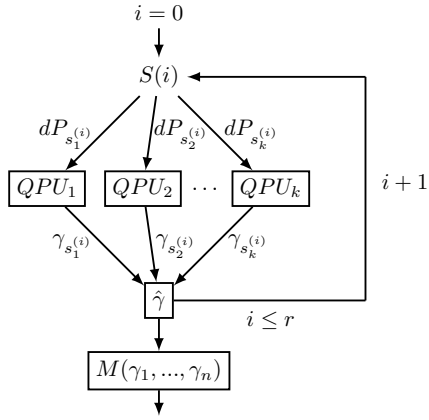

To generate $\mathcal{P}$, the collection of programs and schedule, Algorithm~\ref{algo:sched_and_dist} is used. Input to Algorithm~\ref{algo:sched_and_dist} is 1) The specifications of the distributed quantum computers $Q=[q_1,...,q_n]$; 2) The circuit input to the program with width $w$, that is, the number of qubits simultaneously needed to run the circuit; 3) An algorithm $\mathcal{A}$ which takes $Q$ as input and determines an allocation for $w$ logical qubits or determines no allocation exists; 4) A collection of monolithic programs $\{ P_i \}_{i=1}^n$. The output of the algorithm is a schedule for executing a distributed program $\{\{dP_i\}_{i=1}^{n}, S(i), M\}$. In Fig.~\ref{fig:parallel} is a depiction of how such a system could perform. 

\begin{example}
	Let $\{P_1, ..., P_{10}\}$ be a collection of programs that run circuits with width $w=4$ and $Q=[10, 10]$. If $\mathcal{A}$ is an algorithm that greedily allocates qubits, then the output of Algorithm~\ref{algo:sched_and_dist} is: $S(0) = \{ \{1,2,3\}, \{3,4,5\} \}$, $S(1)= \{ \{6,7,8\}, \{8,9,10\} \}$ and $\{dP_1, ..., dP_{10} \}$, where $dP_3$ and $dP_8$ are distributed between the two QPUs and the other programs run monolithically.
\end{example}

\begin{algorithm}
	\textbf{Input:} QPUs $Q=[q_1, q_2,...,q_k]$, $w$ the circuit width, qubit allocation algorithm $\mathcal{A}$, programs $\{P_1, ..., P_n\}$. Assume $\forall i\leq k, w \leq q_i$.\\
	\textbf{Output}: $\mathcal{P} = \{\{dP_1, ..., dP_n\}, S(i)\}$, $dP_j$ the distributed program for circuit execution $j$, $S(i)$ the schedule for $r$ rounds.
	\begin{algorithmic}[1]
		\State $a\gets 0; i\gets0; dP\gets\{\}; A\gets\{\};$
		\For{circuit $c \leq n$}
		\State Allocate $w$ qubits within current $Q$ with $\mathcal{A}$
		\If{an allocation exists}
		\State $A\Leftarrow$ allocation  \Comment{Append the allocation to $A$}
		\State reduce the available qubits in $Q$ based on allocation
		\State $a \gets a+1$
		\ElsIf{no allocation exists or $c=n$}
		\State Use allocations $A$ to distribute $a$ circuits  \cite[Alg. 3]{diadamo2020vqe}
		\State $dP \Leftarrow$ Generate $a$ distributed programs \cite[Alg. 7]{diadamo2020vqe}
		\State $S(i) \gets \{c-a, ..., c\}$
		\State Reset $Q$; $A\gets\{\}$; $a\gets0$; $i\gets i+1;$
		\EndIf
		\EndFor
	\end{algorithmic}
	\caption{Distributed Quantum Algorithm Scheduler}
	\label{algo:sched_and_dist}
\end{algorithm}

Iterative quantum algorithms mapped to this model and scheduled using  Algorithm~\ref{algo:sched_and_dist} stand to face the same ``horizontal speedup" as mentioned---a run-time speedup achieved by allocating more quantum processors to run in parallel. Influencing this speedup is the algorithm used to allocate qubits---$\mathcal{A}$ in Algorithm~\ref{algo:sched_and_dist}---for distributed processing. The choice of algorithm that solves this problem can come in a variety of flavors. For example, an allocation algorithm that simply chooses qubit allocations randomly will likely introduce more non-local gates, potentially diminishing potential speedups due to the needed additional logic, whereas one which considers the topology and connectivity of the quantum processor can minimize the number of non-local gates. Alternatively, \cite{tang2021cutqc} addresses qubit allocation as to reduce the circuit width in a circuit using a technique called \enquote{circuit cutting} to run parts of a circuit independently and then uses classical post-processing to combine outputs. Their algorithm further aims to minimize the classical post-processing overhead. One can use the technique to define parallel programs to then execute the overall circuit over a cluster of QPUs.

With an optimal allocation algorithm, the speedup of the parallelization for the algorithms we investigate is not found by a reduction in algorithm complexity, but from running multiple iterations of an algorithm simultaneously reducing the run-time of execution. This type of speedup is commonly defined as the ratio between the run-time of one process running an algorithm and the run-time of $p$ parallel processes running a parallelized version \cite{quinn1994parallel}. We also note that in classical distributed computing, the concept known as Amdahl's law is used to predict the theoretical speedup  \cite{amdahl1967validity, hill2008amdahl}. The law predicts that eventually the communication latency between many processors will diminish the reduced runtime of parallel processing. Indeed this  applies to distributed quantum computing as well, but still there are advantages to be gained as analyzed in \cite{beals2013efficient}. 

Future work will require a deeper investigation into how much of a horizontal speedup can be gained in the purely quantum setting and which parameters influence the speedup. One of the parameters that will play a large role will be---as with classical distributed computing---the topology of the network, but indeed there are parameters that will exist only in the quantum setting for distributing algorithms. We plan to further investigate the affects of two such parameters: 1) The quality of created entanglement and 2) Entanglement distribution protocols; each of which will affect the performance of non-local control gates. If the entanglement generation rate is low---which could be the case when using deterministic entanglement generation (roughly in the $\sim100~ms$ regime \cite{humphreys2018deterministic})---then there could be long waiting times during execution. Moreover, when the quality of the entanglement is low (but high enough to be useful), many repetitions of the algorithm could be needed to produce meaningful results. A full investigation will be necessary especially for developing optimized distributed circuit compilation algorithms.

\section{Parallel and Distributed Quantum Algorithms}\label{sec:para-algos}

In this section we describe some examples of quantum algorithms that can be mapped to the parallelized model from the previous section, hence can benefit from horizontal speedup. The property that each of the following algorithms has in common is the quantum part of the algorithm can be split up across multiple QPUs to run in parallel and the classical outputs can then be merged to produce the same result as if the quantum part was instead ran iteratively. The types of algorithms use techniques like output counting or have linearly properties that distribute straightforwardly. We investigate some such examples and explore some of the expected benefits and disadvantages.

\subsection{Variational Quantum Eigensolver}

Computing the eigenvalues of certain quantum operators can be challenging for classical computers due to the exponential scaling in the dimensions of the operators with the increase in the number of quantum states of the system. QPE allows one to compute such eigenvalues in a much more efficient manner, but requires a coherent fully-connected quantum computer to produce good estimates. Consequently, the Variational Quantum Eigensolver~\cite{peruzzo2014variational} (VQE) was proposed as a low-depth alternative, using a hybrid model containing classical optimization and quantum computing. As per the name, VQE belongs to the family of Variational Quantum Algorithms (VQA)~\cite{cerezo2020variational}, a group of hybrid algorithms that include a quantum circuit as a subroutine. VQE uses a classical computer to fine-tune the parameters of the ``ansatz" preparation circuit. In VQE, by tuning the parameters to the ansatz circuit one can minimize an expectation value and use this as an estimate for the minimum eigenvalue. The algorithm is built on the fact that certain Hamiltonian operators can be decomposed into a polynomial number of terms of simpler Pauli operators. As a result, the evaluation of the expectation value of such Hamiltonians reduces to a linear combination of the expectation values of these simpler operators. With this, one can simply measure the different qubits as per the observables in each term to obtain the term's expectation value in constant time and then recombine to find an overall estimate. 

In its standard form, VQE performs this in an iterative fashion, but it can benefit from using a cluster of quantum processors in two different ways. The general workflow would be to dispatch some terms as well as the respective parameters to each quantum processor, which would then compute the expectation value of the terms, and then the dispatching node would aggregate the results from each of the different processors. After carrying out the classical optimization step generating the new parameters for the ansatz, the dispatcher would then the new parameters to the quantum processors to repeat the process. The second advantage comes from the fact that the Hamiltonian governing a molecule requires more quantum systems to simulate as the complexity of the molecule increases. By using a cluster of interlinked quantum computers, one can simulate larger Hamiltonians using the interlinked smaller quantum computers. Indeed with an approximately 48 qubit Hamiltonian, it is predicted to be infeasible for a classical computer to simulate \cite{arute2020hartree}, which could achievable with an interlinked cluster of existing quantum computers.

We now frame this algorithm in the setting of the previous section. For a Hamiltonian $H=\sum_{i=1}^n c_i L_i$, where each $c_i\in\mathbb{R}$ and $L_i \in \{I, \sigma_x, \sigma_y,\sigma_z\}^{\otimes w}$ is a tensor product of $w$ Pauli matrices (or identity), we can form a collection of programs $\{P_i\}_{i=1}^n$ where each $P_i$ is the combined $w$ width ansatz preparation circuit, generating $\ket{\psi_j}$ for the $j$th iteration, prepended before the respective circuit for $L_i$ with $N$ repetitions. The merging function $M$ is simply to reassemble the linear combination, where $M$ applies the respective the coefficient $c_i$, computing the estimate for $\sum_{i =1}^n c_i \langle \psi_j| L_i | \psi_j\rangle$. This type of paralellization has been investigated in depth in \cite{self2021variational}, showing up to a 100-fold improvement in algorithm execution efficiency in experiment in comparison to iterative methods. Indeed many variational quantum algorithms have this same structure \cite{self2021variational, cerezo2020variational} and can be parallelized in a similar fashion, making it more feasible for executing the class of algorithms on near-term devices.

\subsection{Low-Depth Quantum Amplitude Estimation}

Already considered in 2002 by Bassard et al.~\cite{brassard2002quantum}, Quantum Amplitude Estimation (QAE) remains one of the fundamental algorithms for quantum computing, as it adds, for one, a significant performance speed-up for Monte-Carlo methods~\cite{montanaro2015quantum}. An issue to overcome in order to use QAE with near-term quantum computers is to greatly limit the circuit depth. In its original form, QAE uses a combination of QPE and Grover's search~\cite{grover1996fast}, where QPE, with no additional assumptions, uses circuits that deepen proportionally to the inverse of the precision \cite{brassard2002quantum}. Moreover, QPE requires an application of the inverse-QFT algorithm requiring a high-depth and highly-connected quantum processor. To overcome these issues, proposals for low-depth, QFT-free QAE have been proposed~\cite{suzuki2020amplitude,giurgica2020low,grinko2021iterative}. 

From these approaches, we focus on the algorithm called the ``Power Law Amplitude Estimation" (PLAE) algorithm proposed in~\cite[Algorithm 2.1]{giurgica2020low}. PLAE works by using a maximum likelihood estimation routine where for each number of queries $m_k \in \mathcal{K} \equiv \{\lfloor{k^{(1-\beta)/2\beta}}\rfloor: k \leq K \}$---with $K$ bounded above by a constant that grows depending on the desired accuracy and $\beta\in(0,1]$---a circuit making $m_k$ sequential oracle calls is executed on a quantum computer. For each $m_k$, the circuit making $m_k$ queries executes $N$ times, measuring the output of a single qubit, essentially performing tomography. Once all of the $K$ circuits execute, a Bayesian update step is performed iteratively on the $K$ statistics outputs. 
	
In the framework of the previous section, there is a clear parallelization to make for this problem. We can define a program $P_k$ for each $k\leq K$ to be the oracle circuit of width $w$ with $m_k$ oracle queries and $N$ repetitions. The output $\gamma_k$ of $P_k$ is the accumulated statistics of performing $m_k$ oracle queries. The Bayesian update task is used for the merging function $M$. Once all $\{\gamma_k\}_{k=1}^K$ are collected, a phase estimate is made based on the original algorithm. In this way, one can split the load of executing the $K$ circuits across multiple quantum computers, thereby gaining a horizontal speedup. A further parallelization that can be made is to duplicate programs $P_k$ on multiple QPUs, using the same oracle query but dividing the number of circuit repetitions across the QPUs to then merge the counting statistics for each oracle type. Algorithms using Bayesian update methods via counting as with this version of QAE have been proposed in other modified quantum algorithms \cite{wang2019accelerated}, and further investigation for this algorithm class could prove fruitful.  
	
\subsection{Quantum $k$-Means Clustering}
	
	Clustering data into groups based on the properties of the data can be used to find correlations between the data features. $k$-Means clustering is an unsupervised machine learning algorithms used to perform such clustering~\cite{lloyd1982least}. The $k$-Means algorithm takes as input a collection of unlabeled data, or feature vectors, and outputs $k$ clusters, where in each cluster are the data points that minimize the distance to a computed centroid point. The algorithm runs for a number of iterations, improving the centroid locations in order to minimize the average distance between the points in the cluster at each step. A distance metric is used to determine how far apart two data points are from each other. Classically, the usual method for measuring the distance is to simply take the Euclidean distance. For feature vectors of length $N$, computing Euclidean distance requires $O(N)$ steps. Using the quantum encoding known as amplitude encoding, one can encode $N$ length vectors into $O(\log_2 N)$ qubits, an exponential decrease for encoding, assuming one can load quantum states into a quantum random access memory~\cite{lloyd2013quantum}. With this encoding, one can perform a swap test to compute an estimate for the Euclidean distance between two feature vectors. The swap test performs proportionally to the number of qubits used in the encoding, and can lead to---in theory---an exponential decrease in the number of operations used to compute distance. Quantum $k$-means clustering is especially interesting as it is suitable for near-term quantum devices~\cite{khan2019k, johri2021nearest}.
	
	Because each feature vector is compared to each of the $k$ clusters based on the algorithm of~\cite{lloyd1982least}, $n$ distance estimates are made for each of the $k$ centroids. To parallelize this we can consider programs $\{P_{ij}\}_{i=1,j=1}^{n,k}$ where each program computes the distance between feature vector $i$ and centroid $j$. The circuit for each $P_{ij}$ is the one described in~\cite{kopczyk2018quantum}, which loads two feature vectors using amplitude encoding and an additional ancilla qubit for performing the swap test. The merging function $M$ collects the outputs of $i\cdot j$ programs grouping the circuit outputs in $i$ vectors of length $j$ such that the closest centroid can be determined. With this, one can then update the centroid positions classically and repeat the process until convergence is reached, or a maximum number of iterations are performed. 
	
	For a purely parallel version of $k$-Means clustering, the horizontal speedup will scale linearly according to the number of quantum processors until the scale of connectivity comes into play according to Amdahl's law. When moving to the distributed setting, where the number of features cannot be encoded in a single QPU, it becomes important for determining the overall run time to consider how the classical data is encoded in the quantum computer. Indeed depending on how one performs encoding it could be that no quantum advantage is achieved for clustering \cite{tang2021quantum}. If a standard quantum state preparation algorithm is used to perform amplitude encoding across a distributed system of quantum computers, then an exponential number of control gates will be used for state preparation with the number of features, but only a logarithmic number of control gates for performing the swap-test. Alternatively with an angle encoding, only a linear number of control gates are needed for state preparation, but also a linear number of control gates for the swap test, hence no quantum advantage. Moreover, the more control gates needed across a distributed system will result in more classical communication and entanglement generation. The full affect of quantum state preparation across distributed systems will be of interest for future work, especially adapting novel preparation methods as in \cite{araujo2021divide} for distributed systems.

\section{The Interlin-q Simulation Framework}
	
	In this section, we introduce our novel distributed quantum algorithm simulation framework Interlin-q. Interlin-q is a Python framework allowing the simulation of networked quantum computers executing a quantum algorithm distributed over a user-specified topology. The goal of Interlin-q is not to perform high-performance computing, but rather to test and verify the necessary steps of distributing circuits and generating control instructions. The tool is meant for validating these tasks by executing them in this simulated environment and collecting the various statistics regarding the quantity of resources. We begin by summarizing the architecture of Interlin-q and explain its inner-workings. In the next section review demonstrations.

	\subsection{Design Principles}
	
	\begin{figure}
		\centering
		\begin{tikzpicture}[scale=0.75, every node/.style={transform shape}]
			\node at (0, -1) {(a)};
			\node at (2.5, -1) {(b)};
			\node at (5, -2.5) {(c)};
					        
			\node[client] (client) at (0, 0) {};
			\node[server, minimum size=17mm, ] (controller) at (2.5, 0) {};
					    
			\node[nuc, minimum size=10mm] (q1) at (5, 1.5) {};
			\node[nuc, minimum size=10mm] (q2) at (6.75,0) {};
			\node[nuc, minimum size=10mm] (q3) at (5, -1.5) {};
					    
			\node[circle, draw, fill, inner sep=0pt, minimum size=4pt] (dot) at (5, 0) {};
					    
			\draw[<->, line width=.2mm] (client.east) -- (controller.west);
			\draw[<->, line width=.2mm] (dot.west) -- (controller.east);
					    
			\draw[<->, line width=.2mm] (dot.east) -- (q2.west);
			\draw[<->, line width=.2mm] (dot.north) -- (q1.south);
			\draw[<->, line width=.2mm] (dot.south) -- (q3.north);
					    
			\draw[decorate, decoration={snake, amplitude=0.45mm, segment length={2mm}},line width=.2mm] (q1.south east) -- (q2.north west);
			\draw[decorate, decoration={snake, amplitude=0.45mm, segment length={2mm}},line width=.2mm] (q2.south west) -- (q3.north east);
					    
		\end{tikzpicture}
		\caption{The architecture used in Interlin-q: The client (a) constructs a circuit input designed for a monolithic quantum computer and sends it to the controller (b). The controller remaps the circuit for the pre-defined but arbitrary distributed system and generated the execution schedule. Once complete, the respective control instructions are to the quantum computers in (c). The quantum computers execute the schedule and send the results back to the controller who processes the results to send back to the client.}
		\label{fig:architecture}
	\end{figure}
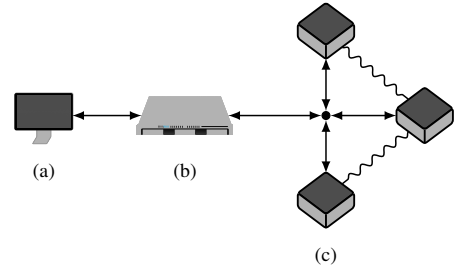
	
	The simulated architecture of Interlin-q consists of three types of network nodes: The client node, the controller node, and the computing node. In Fig. \ref{fig:architecture} is a depiction of the assumed architecture of Interlin-q. We base the design principles of Interlin-q on these node types, as a distributed quantum system will likely follow such an architecture. The responsibilities of each node type are as follows:
	
	\textbf{Client Node}: The client is a user terminal node where the user of the system inputs the program information in a similar way as described in the previous section. A monolithic circuit is specified along with the merging function. This information is then passed forward to the controller node which continues the execution process. Once execution is complete, the controller node returns the results of the program after performing the merging function to the client node. This isolation of the client node removes the need for the user of the system to know the underlying architecture of the distributed quantum computer.
	
	\textbf{Controller Node}: The controller node is the conductor which orchestrates the distributed system of quantum computers. It is aware of the distributed topology and the quantum processor architecture of each node in the network and therefore can define allocations for qubits for circuit execution. Once the program information from the client node is provided, the controller node can perform Alg. \ref{algo:sched_and_dist} to prepare for execution and awaits outputs from each program. Once outputs are received, the controller merges the results accordingly and responds to the client node.
	
	\textbf{Computing Nodes}: The computing nodes execute quantum algorithms based on the instructions provided by the controller node. A computing node has the ability to prepare and perform logical operations on qubits, store qubits and shared EPR pairs, and also perform any classical post-processing. Computing nodes further can communicate with other computing nodes in the network to share EPR pairs or transmit classical information. Computing nodes are networked via both a classical network for transmitting purely classical data and a quantum network for generating entanglement amongst themselves. Networked computing nodes also share a synchronized clock in order to maintain synchronization, important for two qubit operations and joint measurements. Once execution is complete, the measurement results are sent back to controller node.
	
	\subsection{Simulated Setup and Preprocessing}
	
	To implement these design principles, Interlin-q has Python classes to represent the controller and client nodes specified in the previous subsection where the client node is assumed the to be the user of Interlin-q. Further, abstractions of circuits are developed for automatic circuit remapping. To create the simulated distributed computing environment, the user firstly initializes the computing nodes and a controller node. The network is initialized and configured using QuNetSim, thereby defining network nodes and topology. In its current state, Interlin-q will assume that the computing node form a complete network and that each is connected to the controller node. Because of this, Interlin-q has a built-in function for generating the network where future work involves allowing for various network topologies.
	
	To initialize a computing node using the respective class, one specifies the number of qubits and optionally the duration of the various quantum gates, or gate times, for the gates that computing node supports. The latter is especially useful when connecting a network of quantum computers each realized using different qubit technologies, for example connecting one QPU based on superconducting with one using trapped ions, where the two technologies differing greatly in gate times. With known gate times, precise algorithm execution schedules can be generated \cite{diadamo2020vqe}. Within Interlin-q is a simulated synchronized clock. Each of the computing nodes share a singleton clock object and function such that for each tick of the clock, depending on the execution time of their instruction, perform a specific operation. The set of instructions is generated dependent on the gate times. The computing nodes and the nodes and their qubits are assigned unique IDs to be used for circuit creation.
	
	A controller node is also initialized by the user and the collection of computing nodes is passed as an initialization parameter to the controller. To create a simulation, a user generates a monolithic circuit as a parameter to give to the controller where built-in to Interlin-q is a feature to remap it to the distributed computing nodes. In the current state of development of Interlin-q, a custom circuit class is used to both enable the distribution of the circuit but also to simplify user input process. The circuit model we used is composed of qubits and layers, the standard circuit model for quantum computing. To create a circuit in Interlin-q, a user, acting as the client node, specifies the gates for circuit qubit by qubit. Once the circuit object is created, it is passed to the controller and then automatically distributed based on the network of computing nodes. Built-in to the controller node is the conversion of the circuit model to a layered model, where a new layer is created for each $i\leq w$ where $w$ is the longest sequence of gates in the circuit. With the layering generated, the controller performs an algorithm equivalent to \cite[Alg.~3]{diadamo2020vqe}. To summarize the referenced algorithm, Interlin-q processes the circuit layers one by one and determines if any of the gates are distributed across distinct QPUs. If found, the necessary logic to create an equivalent distributed circuit is filled. This is done repeatedly until all non-local control gates are generated, forming an equivalent distributed circuit.
	
	With the now remapped circuit, the controller node generates an execution schedule, creating the instructions to distribute to send to each computing node. To generate the list of control operation, an equivalent algorithm to \cite[Alg.~7]{diadamo2020vqe} is implemented. For each gate in the remapped circuit, the algorithm maps it to a logical instruction, including any gate parameters, and also marks the control instruction with a timestamp. In practice, the networked QPUs would use a reference clock to execute instructions to maintain the needed synchrony for distributed instructions. This logic is simulated in Interlin-q using a custom shared clock class. The collection of control instructions are given an integer timestamp so that during execution the multi-threaded simulation performs according to the control instruction order. To complete the preprocessing stage, the controller completes the generation of the execution schedules according with Alg.~\ref{algo:sched_and_dist} and transmits a broadcast message to the computing nodes.
	
	\subsection{Execution}
	
	Execution begins when the controller node broadcasts the instruction sets to the computation nodes. Because Interlin-q is built on QuNetSim, the simulated environment runs multi-threaded, each node in the network running in its own thread. The computation nodes therefore await control messages to perform their tasks. Once the specific scheduling message is received by the computing nodes, they begin carrying out the instructions chronologically depending on the timestamp of the instructions. An instruction is either completely local or can be non-local. A non-local instruction is performed according to Fig. \ref{fig:cat-entangler}, where entanglement is generated between two computing nodes and used to transmit control information. The schedule will be such that at the same time instance one node will wait for an EPR pair to arrive followed by a classical bit, performing their part of the cat-entangler while the other will send the EPR pair, measure their half and send the results onward. Once the instructions are all carried out by each computing node, the time moves forward by one unit. In reality, this clock will be independent of computing nodes, and gate times will known such that precise time schedules can be generated. When the set of instructions is completed, the computing nodes transmit their measurement outcomes back to the controller and receive a new set of instructions if there are more to receive, otherwise the controller proceeds to merge the outputs and can output the results to the simulation.
	
	\subsection{Related Platforms}
	
	Related to this project are platforms that use batched circuit execution in a parallelized system. As far as we know, the only such example publicly available is part of the Qiskit Runtime Services offered by IBM called \enquote{circuit-runner}~\cite{ibm-circuit}. The circuit-runner service takes as input a collection of un-optimized, pre-compiled quantum circuits and sends them to the IBM cloud service to be executed on their network of quantum computers and simulators. Once arrived, the circuits are optimized and complied online, and executed on the selected hardware backend. Once execution is complete the output information for the circuits such as measurement results, duration, and more is sent back to the user. This sequence of steps for executing batch circuits is much like in the steps for how one uses Interlin-q. Indeed, behind the scenes, IBM's circuit-runner service could be using distributed quantum computing when it becomes available and run much like the structure Interlin-q is built on. In terms of using circuit-runner for running parallelized quantum circuits, one could in fact make use of the paradigms introduced in Section \ref{sec:para-algos}. Where Interlin-q differs is that it opens the \enquote{black-box} into how the quantum computers are interacting behind the scenes. By adding a simulated quantum network, a user can investigate precisely how a networked quantum computer executes a distributed quantum program.  
	
	Interlin-q is build on top of the quantum network simulator QuNetSim \cite{diadamo2020qunetsim} mainly due to its real time simulation design. Indeed other quantum network simulators exist, but are built more towards simulating the hardware properties of quantum networks rather than a focus on application development. Other than QuNetSim, another viable quantum network simulation platform is SimulaQron \cite{dahlberg2018simulaqron} which also runs in real time, the key differences which are detailed in \cite{diadamo2020qunetsim}. We focus on real time simulators because of the key design principle of Interlin-q which is the shared clock. The shared clock controls the execution of each thread executing instructions, which is more aligned with a distributed quantum system, rather than knowing the execution schedule ahead of time as discrete-event simulators do, such as in NetSquid~\cite{coopmans2021netsquid} and SeQUeNCe~\cite{wu2021sequence}. Indeed, using such discrete-event based simulators for benchmarking would be a valuable extension of Interlin-q, but our current focus is on verification of distributed quantum algorithms and estimating their resources.
	
	\section{Demonstrations}
	
	To demonstrate the current abilities of Interlin-q, in this section we review some demonstrations and the full source code of the project can be found at~\cite{interlinqgit}. To begin, we will go through an example of distributed quantum phase estimation making use of Interlin-q's built in distribution mapping. Next, we review an example of a parallelized version of VQE.
	
	\subsection{Distributed Quantum Phase Estimation}
	
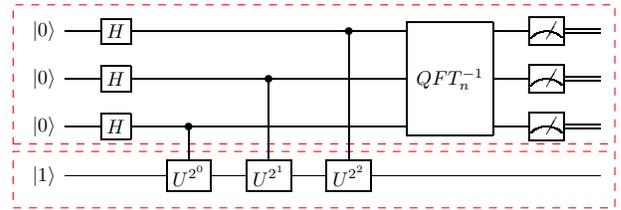
\begin{figure}[ht]
		\centering
		\begin{tikzpicture}
			\draw[draw=red, dashed] (-4.25, 1.325) rectangle ++(8, -1.85);
			\draw[draw=red, dashed] (-4.25, -.625) rectangle ++(8, -.75);
			\node[scale=0.75] at (0, 0) {
				\Qcircuit @C=1.8em @R=1.0em {
					\lstick{\ket{0}} & \gate{H} & \qw & \qw & \ctrl{3} &\multigate{2}{QFT_n^{-1}} &\meter & \cw   \\  
					\lstick{\ket{0}} & \gate{H} & \qw & \ctrl{2} & \qw & \ghost{QFT_n^{-1}} &\meter & \cw \\
					\lstick{\ket{0}} & \gate{H} & \ctrl{1} & \qw & \qw & \ghost{QFT_n^{-1}} &\meter & \cw\\
					\lstick{\ket{1}} & \qw & \gate{U^{2^0}} & \gate{U^{2^1}} & \gate{U^{2^{2}}} & \qw &  \qw &  \qw
				}
			};
		\end{tikzpicture}
		\caption{Circuit diagram for QPE with unitary operation $U$ for this example.}
		\label{fig:qpe-circuit}
	\end{figure}

	In this example, we demonstrate a version distributed QPE using the circuit configuration as depicted in Fig.~\ref{fig:qpe-circuit}. In this simulated architecture, we place the measurement qubits on one quantum computer (the upper portion of the circuit), and the qubit whose phase to estimate on another (the lower portion). In this case, the control unitary gates, since they are non-local, will need additional instructions added in order to perform them correctly. In this example, we see we just need to build the circuit as it is depicted, and Interlin-q will then perform the circuit remapping and carry out the execution of the instructions.
	
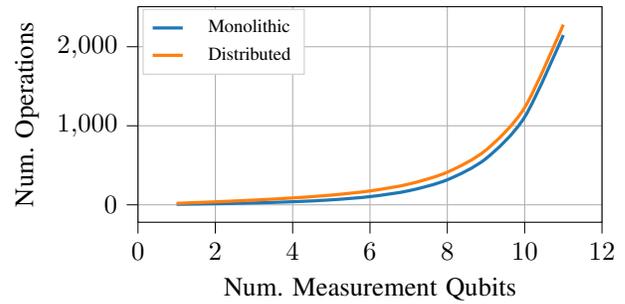
\begin{figure}
	    \centering
\begin{tikzpicture}

\definecolor{color0}{rgb}{0.12156862745098,0.466666666666667,0.705882352941177}
\definecolor{color1}{rgb}{1,0.498039215686275,0.0549019607843137}

\begin{axis}[
tick align=outside,
tick pos=left,
legend cell align={left},
legend style={
  fill opacity=1,
  draw opacity=1,
  text opacity=1,
  font=\scriptsize,
  at={(0.01, 0.99)},
  anchor=north west,
  draw=white!80!black,
},
legend image code/.code={
    \draw[mark repeat=2,mark phase=2]
    plot coordinates {
    (0cm,0cm)
    (0.15cm,0cm)        
    (0.4cm,0cm)         
};%
},
x grid style={white!69.0196078431373!black},
xmajorgrids,
ymajorgrids,
height=4.45cm,
width=7.75cm,
xlabel={Num. Measurement Qubits},
ylabel={Num. Operations},
xtick style={color=black},
y grid style={white!69.0196078431373!black},
ytick style={color=black},
]
\addplot [very thick, color0, smooth]
table {%
1 7
2 14
3 24
4 39
5 63
6 104
7 178
8 317
9 585
10 1110
11 2148
};
\addlegendentry{Monolithic}
\addplot [very thick, color1, smooth]
table {%
1 19
2 38
3 60
4 87
5 123
6 176
7 262
8 413
9 693
10 1230
11 2280
};
\addlegendentry{Distributed}
\end{axis}

\end{tikzpicture}
	    \caption{The number of operations required to perform quantum phase estimation on one qubit with varying level of precision when using a distributed system of two QPUs.}
	    \label{fig:qpe-total-gates}
	\end{figure}

	To start the example, we import the necessary libraries and initialize the configuration:

\begin{lstlisting}
import numpy as np
# QuNetSim Components
from qunetsim.components import Network
from qunetsim.backends import EQSNBackend
# Interlin-q Components
from interlinq import ControllerHost, Circuit, ComputingHost, Constants, Qubit

# Initializing network objects
network = Network.get_instance()
network.start()
controller_host = ControllerHost(
        host_id="controller",
        backend=EQSNBackend()
    )
# Create a network of distributed QPUs
computing_hosts, q_map = controller_host \
  .create_distributed_network(
     num_computing_hosts=2,
     num_qubits_per_host=3
  )
# Start the controller and create the network
controller_host.start()
network.add_host(controller_host)
network.add_hosts(computing_hosts)
\end{lstlisting}
	We make use of QuNetSim's network object and add the controller and computing nodes to the network. The \verb|create_distributed_network| ControllerHost method  will generate a completely connected network topology and in this case specifically, with two ComputingHosts each with three qubits. The next step is to define the protocol logic for each of the network nodes:
\begin{lstlisting}
def computing_host_protocol(host):
    """
    Protocol for the computing host
    """
    host.receive_schedule()
    host.send_results()
    
def controller_host_protocol(host, q_map, input_gate):
    """
    Protocol for the controller host
    """
    # Generate the circuit for QPE
    circuit = qpe_circuit(q_map, input_gate)
    host.generate_and_send_schedules(circuit)
    # Block until measurement results arrive
    host.receive_results()
    meas_results = host.results["QPU_1"]["val"]
    output = [0] * 3
    print(results)
    for qubit in meas_results.keys():
        output[int(qubit[-1])] = meas_results[qubit]
    decimal_value = 0
    output.reverse()
    for i, bit in enumerate(output):
        decimal_value += ((2 ** i) * bit)
    phase = decimal_value / 8
    print("The estimated phase is {0}".format(phase))
\end{lstlisting}
	The actions of the computing host generally have the same structure which is to await their instructions from the controller and then to send the measurement results back to the controller. The controller on the other hand takes as input the network topology (assumed to be completely connected) and a unitary in which to use for the phase estimation step. The controller uses the information to then generate the circuit, generate control instructions, and then sends it to the computing nodes, awaiting the measurement results to perform the post-processing step. To generate the circuit:
\begin{lstlisting}
def phase_gate(theta):
    return np.array([[1, 0], [0, np.exp(1j * theta)]])
    
def quantum_phase_estimation_circuit(q_map, client_input_gate):
    """
    Returns the monolithic circuit for quantum phase estimation algorithm
    """
    phase_qubit = Qubit(computing_host_id='QPU_0', q_id=q_map['QPU_0'][0])
    phase_qubit.single(Operation.X)
    meas_qubits = []
    
    for q_id in q_map['QPU_1']:
        q = Qubit(computing_host_id='QPU_1', q_id=q_id)
        q.single(Operation.H)
        meas_qubits.append(q)
    for i, q in enumerate(meas_qubits):
        for _ in range(2 ** i):
            q.two_qubit(Operation.CUSTOM_CONTROLLED, phase_qubit, client_input_gate)
    # Inverse Fourier Transform
    meas_qubits.reverse()
    for i, q in enumerate(meas_qubits):
        for j, q2 in enumerate(meas_qubits[:i]):
            q2.two_qubit(Operation.CUSTOM_CONTROLLED, q, phase_gate(-np.pi * (2 ** j) / (2 ** i)))
        q.single(gate=Operation.H)
    # Measure the qubits
    for q in meas_qubits:
        q.measure()
    return Circuit(q_map, qubits=meas_qubits + 
                                [phase_qubit])
\end{lstlisting}
	Finally, to begin execution and wait for results:
\begin{lstlisting}
# For phase = 1/3
input_gate = np.array([
                        [1, 0], 
                        [0, np.exp(1j * 2 * np.pi / 3)]
                     ])
t1 = controller_host.run_protocol(
    controller_host_protocol,
    (q_map, input_gate))
computing_hosts[0].run_protocol(computing_host_protocol)
computing_hosts[1].run_protocol(computing_host_protocol)
t1.join()
network.stop(True)
\end{lstlisting}
	Since the backend simulator we selected, EQSN, is noiseless simulator, we need just use one shot to get an estimate of the phase. The output in this case is: \verb|The estimated phase is 0.375| as expected for a 3 bit estimation. In this simulation, we can gather statistics about how much communication is involved in the network to execute the algorithm. For the example, we have used 3 measurement qubits, but by changing the parameter \verb|num_qubits_per_host|, we can adjust the number of measurement qubits. In Fig.~\ref{fig:qpe-total-gates}, we plot the number of total operations needed, including the number of gates, entanglement generation, and classical communication between the nodes to execute the algorithm, comparing a monolithic version to a distributed version when the measurement qubits are separated as in Fig.~\ref{fig:qpe-circuit}. In this case, many control operations can be performed via small amounts of classical communication and entanglement because the control information for each measurement qubit the for non-local gates can be transferred using one EPR pair and two classical messages. 
	
	\subsection{Parallel Ground-State Estimation of $H_2$}
	
	To demonstrate the current parallelization abilities of Interlin-q, we review a demonstration simulation of estimating the ground state energy of the $H_2$ molecule using VQE over a distributed architecture. To get started with Interlin-q, we import the necessary libraries. Interlin-q is built on the QuNetSim~\cite{diadamo2020qunetsim} framework, and uses the QuTiP~\cite{johansson2012qutip} Python backend of QuNetSim for qubit simulations. For this example, we use Xanadu's Pennylane library~\cite{bergholm2018pennylane} for its chemistry features. Here we highlight the structure of the simulation, and the full simulation with all the details is found in the source code~\cite{interlinqgit}. The simulated architecture for this example is the same as depicted in Fig.~\ref{fig:architecture} having three computing nodes.
	
\begin{lstlisting}
# QuNetSim Components
from qunetsim.components import Network
from qunetsim.backends import QuTipBackend
# Interlin-q Components
from interlinq import (ControllerHost, Circuit, ComputingHost, Constants, Clock, Qubit)
# Xanadu's Pennylane Python package
from pennylane import GradientDescentOptimizer
# A wrapper for Pennylane's qchem library
from hamiltonian_decomposition import decompose
\end{lstlisting}
	Once all libraries are imported, we can initialize the network. The network is composed of extended Hosts from QuNetSim for the computing and controller nodes and added to a QuNetSim Network. 
\begin{lstlisting}
# Initializing network objects
network = Network.get_instance()
qutip = QuTipBackend()
network.start(backend=qutip)
clock = Clock()
controller_host = ControllerHost(
        host_id="controller",
        clock=clock,
        backend=qutip
    )
# Create a network of distributed QPUs
# Generates 3 computing nodes each with 4 qubits
# q_map is the distributed network topology metadata 
computing_hosts, q_map = controller_host \
  .create_distributed_network(
     num_computing_hosts=3,
     num_qubits_per_host=4
  )
# Start the controller and create the network
controller_host.start()
network.add_host(controller_host)
network.add_hosts(computing_hosts)
\end{lstlisting}
	We define the parameters for the $H_2$ molecule and generate the Hamiltonian decomposition.
\begin{lstlisting}
geom = "h2.xyz"
charge = 0
mult = 1
basis_set = "sto-3g"
name = "h2"
# Decompose the Hamiltonian to a sum
coefficients, observables, qubit_num = \ 
 decompose(name, geom, charge, mult, basis_set)
# Create tuples for the observable and coefficients 
terms = zip(coefficients, observables)
\end{lstlisting}
	In the case of $H_2$, there are 15 terms in the decomposition and they are distributed evenly to the 3 computing nodes. The controller instructions are broken into two parts. The first part signals the computing nodes to receive the ansatz circuit. Next, the information for each computing node for computing the expectation value terms are distributed evenly amongst the computing nodes, and the instructions for sending back the results are provided. The computing nodes are programmed to simply await instructions and execute.
\begin{lstlisting}
# Protocols for the controller node
def ctrl_preparation_ansatz(node, q_map, params):
  # Define the operations for the ansatz circuit
  ops = init_ops(q_map) + ansatz_ops(q_map, params)
  # Create a single circuit as if one large QPU
  ansatz_circuit = Circuit(q_map, ops)
  # Distribute the circuit and sends execution schedules
  node.generate_and_send_schedules(ansatz_circuit)
  
def ctrl_expectation_values(node, q_map, terms):
  # Prepare the expected expectation value 
  node.schedule_expectation_terms(terms, q_map)
  # Generate Pauli string circuits
  expvals_compute = hamilton_ops(node, q_map)
  # Request expectation values
  expvals_collect = request_expvals_ops(q_map)
  # Create single circuit
  circ = Circuit(q_map, expvals_compute + expvals_collect)
  # Distribute the circuit and sends execution schedules
  node.generate_and_send_schedules(circ)
  # Await results from computing nodes
  node.receive_results()

# Protocol for computing nodes
def cpu_task(node, send_exp):
  node.receive_schedule()
  if send_exp:
    node.send_results("expectation")
\end{lstlisting}
	To execute the simulation for one parameterized ansatz, a series of threads are created so all instructions can perform in parallel. For each part of the instructions, the termination of each thread is monitored before beginning the next part of instructions.
\begin{lstlisting}
# Trigger the distribution of the ansatz circuit
t1 = controller_host.run_protocol(ctrl_preparation_ansatz, (q_map, params))
threads = [t1]
for host in computing_hosts:
  # Trigger the retrieval of the ansatz circuit    
  threads.append(host.run_protocol(cpu_task, (False,))
# Await all threads to terminate
for thread in threads:
  thread.join()
# Trigger the expectation value calculation instructions  
t1 = controller_host.run_protocol(ctrl_expectation_values,
    (q_map, terms))
threads = [t1]
for host in computing_hosts:
  # Await the expectation value calculation instructions    
  threads.append(host.run_protocol(cpu_task, (True,))
# Await all threads to terminate
for thread in threads:
  thread.join()
# Expectation value results ready
print(controller_host.results)
\end{lstlisting}
	To reduce communication resources for this example, the computing hosts are programmed to compute the partial summation of their assigned terms multiplied by the respective coefficient. To then collect the expectation value estimates, the computing nodes send their partial summations to the controller node who receives:
\begin{lstlisting}
{
 "QPU_0": {..., "out": 0.32072},
 "QPU_1": {..., "out": -0.15644},
 "QPU_2": {..., "out": -0.36714}
}
\end{lstlisting}
	The final step is for the controller to sum the three outputs generating an energy estimate for one set of parameters for the ansatz. To complete the optimization procedure, we use the simulation with a Python optimization library, where the variable \verb|params| is the variable to optimize over.
	
	\section{Conclusions and Outlook}
	
	In summary, we formalized---in a mathematical sense---parallel and distributed quantum algorithms and provided a high-level algorithm for scheduling parallel and distributed programs in a network of distributed quantum processors. Within this formalism we analyzed three quantum algorithms that have a structure that maps straightforwardly into the framework we proposed. Lastly, we introduced the Interlin-q software platform for distributed quantum algorithms, a first of its kind. We described its current features and reviewed two demonstrations for how one can use it to simulate distributed and parallelized quantum algorithms.
	
	From a theoretical view, further research directions coming from this work will be to compare the theory of classical parallelized computing to quantum parallelized computing in depth. In this regard, studying the advantages and disadvantages of quantum parallel computing will be highly important, especially as scaling quantum computers via quantum networks becomes more viable. In the development view, for the Interlin-q platform, we aim to further develop its features, implementing more varieties of non-local control to better compare them. As we study the parallelization of quantum algorithms further, it will become important to ensure Interlin-q remains easy to use but also has all of the features needed to simulate various classes of parallelized quantum algorithms. Adding features to easily implement the different ways to parallelize quantum algorithms will be a priority. 
	
	Lastly, the overall goal of the project is to develop Interlin-q into a control platform for distributed quantum hardware. Much like IBM's circuit-runner service, we want to allow users to input their various circuits to execute as a batch-job, but further to be able to control the way their algorithms are distributed, using the various current and forthcoming methods, to then be ran on a distributed quantum computer. With the design methodologies used to develop the platform, as well as the feature of interchangeable quantum backends of QuNetSim, the components of Interlin-q are well architected to be extended into a control plane. The first step to achieving this goal will be to develop a proof-of-concept distributed simulation over a classical network.
	
	Overall, exploring parallel and distributed quantum computing via theory and simulation as this work begins to do will be an important step on the path towards large-scale quantum computing.

	\section*{Acknowledgments}
	
	This work was completed under the Quantum Open Source Foundation Mentorship program and all authors thank the organizers. R.P. and S.D. thank the Unitary Fund for supporting and funding this project. A.R. and S.D. thank Will Zeng for discussions regarding parallelization of QAE. This work is partially funded by the Deutsche Forschungsgemeinschaft (DFG) the Emmy-Noether grant NO-1129/2 (S.D).



\end{document}